\documentclass[fleqn,10pt]{wlscirep}
\usepackage[utf8]{inputenc}
\usepackage[T1]{fontenc}
\usepackage{graphicx}

\newcommand{\og}{\color{black}}
\newcommand{\kf}{\color{black}}

\usepackage{graphicx}
\usepackage{epsfig}
\usepackage{dcolumn}
\usepackage{bm}
\usepackage[utf8]{inputenc}
\usepackage[T1]{fontenc}
\usepackage{color}
\newcommand     {\beq}[1]         { \begin{equation} #1 \end{equation} }
\newcommand     {\beqa}[1]        { \begin{eqnarray} #1 \end{eqnarray} }

\title{Scaling laws of failure dynamics on complex networks}

\author[1]{Gerg\H o P\'al}
\author[1]{Zsuzsa Danku}
\author[1]{Attia Batool}
\author[1]{Vikt\'oria K\'ad\'ar}
\author[2]{Naoki Yoshioka}
\author[2]{Nobuyasu Ito}
\author[3]{G\'eza \'Odor}
\author[1,4,*]{Ferenc Kun}

\affil[1]{Department of Theoretical Physics, Doctoral School of Physics, 
Faculty of Science and Technology, University of Debrecen, H-4002 Debrecen, P.O.Box: 400, Hungary}
\affil[2]{RIKEN Center for Computational Science, 7-1-26 Minatojima-minami-machi, Chuo-ku, Kobe, Hyogo 650-0047, Japan}
\affil[3]{Centre for Energy Research, Institute of Technical Physics and Materials Science, P.O. Box 49, H-1525 Budapest, Hungary}
\affil[4]{Institute for Nuclear Research (Atomki), P.O. Box 51, H-4001 Debrecen, Hungary}
\affil[*]{Corresponding author: ferenc.kun@science.unideb.hu}

\begin{abstract}
The topology of the network of load transmitting connections plays an essential role in the cascading failure 
dynamics of complex systems driven by the redistribution of load after local breakdown events. 
In particular, as the network structure is gradually tuned from regular to 
completely random a transition occurs from the localized to mean field behavior of failure spreading. Based on finite size scaling in the fiber bundle model of failure phenomena, here we demonstrate that outside the localized regime, the load bearing capacity and damage tolerance on the macro-scale, and the statistics of clusters of failed nodes on the micro-scale obey scaling laws with exponents which depend on the topology of the load transmission network and on the degree of disorder of the strength of nodes. Most notably, we show that the spatial structure of damage governs the emergence of the localized to mean field transition: as the network gets gradually randomized failed clusters formed on locally regular patches merge through long range links generating a percolation like transition which reduces the load concentration on the network. The results may help to design network structures with an improved robustness against cascading failure.
\end{abstract}

\begin{document}

\flushbottom
\maketitle
\thispagestyle{empty}

\section*{Introduction}
Cascading dynamics where the local activity of an element triggers a sequence of activated events is a generic feature of a large variety of complex systems. Examples can be mentioned from the activity patterns of neural networks \cite{dahmen_brain_aval_prl_2012,lucilla_neural_2015,neural_cascade_2020} to the intermittent spreading of information or diseases in social communities \cite{newman_epidemic_pre2002,barthelemy_spatial_2011,satorras_epidemic_2015}. Failure cascades form a distinct class of cascading activities because as a cascade propagates, elements of the system become irreversibly inactive without any ability to support load again. The gradual reduction of the load bearing capacity together with the constraint of load conservation can easily give rise to large scale breakdown events spanning a macroscopic fraction
of the system. Cascading failure phenomena frequently occur in our technological environment such as the cascading blackouts of high voltage power grids \cite{chaos_blackout_review_2007}, the breakdown of communication and urban traffic networks \cite{satorras_epidemic_2015,newman_epidemic_pre2002,barthelemy_spatial_2011,dobson_ieee_2012,profile_network_2017}or the fracture of heterogeneous materials \cite{salje_pre2002,PhysRevLett.112.115502,danku_PhysRevLett.111.084302,rosti_crackling_2009} often inducing huge economical costs \cite{boccaletti_complex_2006,chaos_blackout_review_2007,Moreno_2002,trpevski_model_2010}. 

Cascading failures are driven by the redistribution of load following local breakdown events, which increases the load and in turn gives rise to a sequence of secondary failures in the vicinity of failed elements. It is a question of high theoretical and practical importance how the interplay of the structure of the underlying network of load transmitting connections and of the stochastic strength of the elements of the system affects the emergence of failure cascades and the overall robustness of the system against cascading breakdown. The fiber bundle model (FBM) as a generic model of cascading failure phenomena has proven indispensable to obtain a deeper insight into the dynamical and statistical aspects of failure spreading \cite{hidalgo_avalanche_2009,hansen2015fiber,bikas_review_frontiersrt_2020}. Originally introduced to study the fracture and breakdown of heterogeneous materials, an FBM consists of a bundle of fibers arranged on a regular lattice. Under a gradually increasing external load, fibers fail and transfer their load to their immediate neighbors along the edges of the lattice. Recently, it has been demonstrated in FBMs that gradually randomizing an initially regular network of load transmitting connections, a transition occurs from 
the localized universality class of failure phenomena, where catastrophic collapse abruptly occurs after a small amount of damage, to the mean field universality class where global breakdown is preceded by a large number of cascades with a scale free statistics \cite{attia_chaossolit_2022,attia_chaos_2022}. The disorder of the strength of the elements of the system turned to have a stabilizing effect on the network in the sense that reducing disorder makes the localized to mean field transition more abrupt \cite{attia_chaossolit_2022,attia_chaos_2022}.
In spite of the theoretical efforts, the size scaling of the ultimate strength and of the damage tolerance of the system, furthermore, the microscopic origin of the localized to mean field transition remained open problems of fundamental importance.

Based on large scale computer simulations here we determine the scaling laws of failure phenomena of complex networks both on the micro- and macro-scales. We perform FBM simulations on load transmission networks generated by randomly rewiring an initially square lattice of fibers. Topological randomness is controlled by varying the rewiring probability which tunes the network structure from regular to completely random at several system sizes. 
Finite size scaling analysis revealed that for network topologies outside the localized regime the overall load bearing capacity and damage tolerance of the system on the macro-scale, furthermore, the statistics and spatial structure of damage clusters on the micro-scale obey generic scaling laws with scaling exponents which depend on the topology of the underlying load transmitting network. Our calculations showed that the spatial structure of damage accumulating as cascades proceed governs the emergence of the localized to mean field transition. Most notably, we demonstrate that the transition occurs as a consequence of a structural transition of failed nodes similar to percolation: as the network of load transmitting connections is gradually randomized, clusters of failed nodes grown on locally regular patches merge into a dominating cluster through long range links, which in turn allows for a gradual reduction of load concentrations. 
The results can help to design network structures with an improved robustness against cascading failures.

\section*{Results}

\subsection*{Fiber bundle model of cascading failure dynamics on complex networks}
Our study is based on the fiber bundle model where the load transmitting connections of fibers are represented by a complex network with tunable structural properties.
The FBM is one of the most important modelling approaches to the cascading failure dynamics, which was originally developed to study the gradual fracturing of heterogeneous materials. However, due to the generality of its cascading mechanism, FBMs have gained applications in diverse fields including the modelling of cascading blackouts of power grids \cite{power_fiber_2015,dou_robustness_2010,domansky_frontiers_fbm2020}, or the breakdown of urban traffic networks \cite{fbm_traffic_ijmpc_2008,talbot_traffic_pre_2015,bikas_physica_2006}, and flow channels \cite{talbot_traffic_pre_2015}. 
\begin{figure}[ht]
\begin{center}
\includegraphics[scale=0.55]{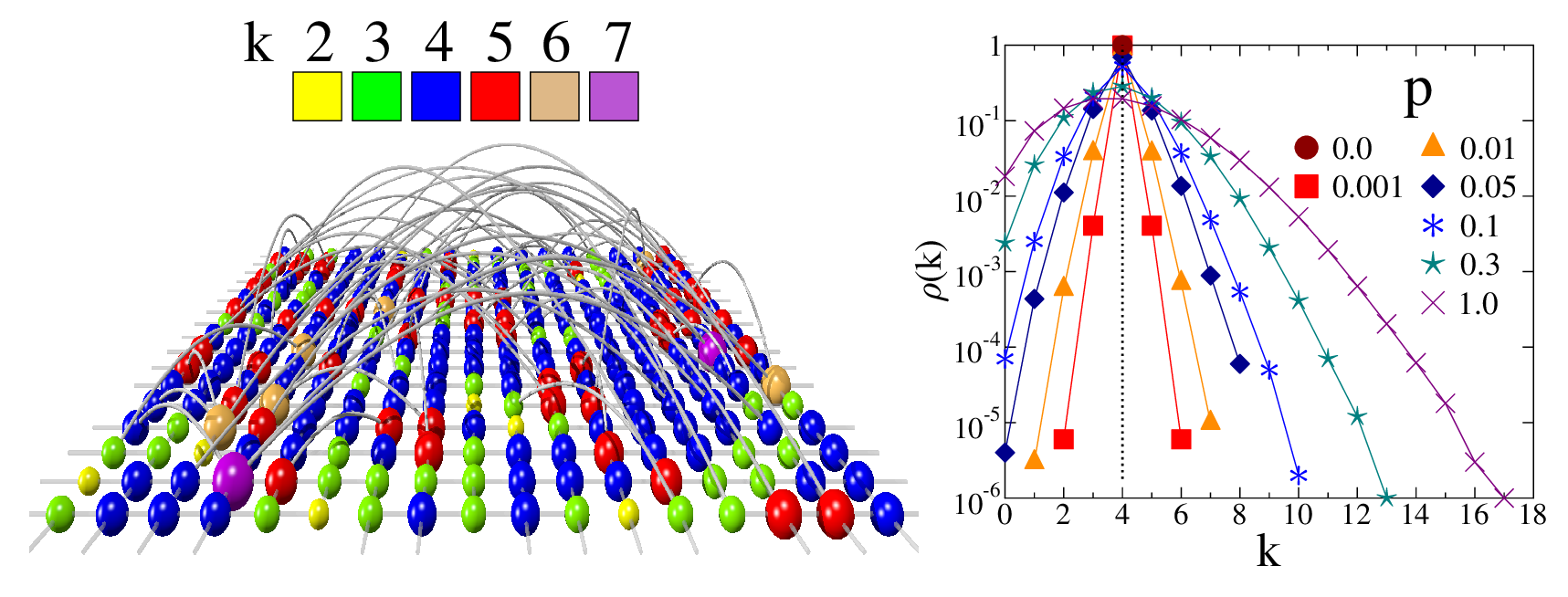}
\caption{Illustration of the model construction. $(left)$ A small lattice of size $L=17$ is rewired at $p=0.1$. The nodes of the network are represented by spheres colored according to their degree $k$. To make the structure of the network transparent the size of the spheres is proportional to the node degree $k$ and the rewired links run outside the plane of the original lattice. $(right)$ The degree distribution $\rho(k)$ of rewired lattices of size $L=1000$ at different rewiring probabilities $p$. At $p=0.0$ the distribution $\rho(k)$ has a single point at $k=4$.}
\label{fig:demo}
\end{center}
\end{figure}
The main advantage of the model is that it allows for a straightforward representation of the fluctuating local strength and of the interaction of the constituents of the system.  
In the model construction we consider a bundle of $N$ parallel fibers 
which are assigned initially to the nodes of a square lattice of side length $L$ so that $N=L^2$ holds. Fibers (nodes) are connected by the edges of the lattice to their four nearest neighbors with periodic boundary conditions, hence, initially the degree $k$ of all nodes is $k=4$. As to the next, each of the $2N$ initially existing connections is rewired according to the Watts-Strogatz {\og (WS)} algorithm \cite{watts_networks_1999}: with probability $p$ each link is removed and then re-established between two randomly selected nodes with the constraint that neither multiple links nor loops are allowed in the system. This rewiring process introduces long range randomized connections, and hence, broadens the degree distribution $\rho(k)$ of the network. Consequently, the topology of the emerging network gradually changes from regular to completely random as the rewiring probability $p$ is varied from $p=0$ to $p=1$. Figure \ref{fig:demo} illustrates the model construction along with the evolution of the distribution $\rho(k)$ of node degree $k$ as the rewiring probability $p$ is increased.

Under a slowly increasing external load $\sigma_0$ the nodes fail when the local load on them $\sigma_i$ exceeds their load bearing capacity $\sigma_{th}^i$ ($i=1,\ldots, N$). The strength of fibers $\sigma_{th}$ is a random variable which is sampled from a probability distribution $p(\sigma_{th})$. To generate the random failure thresholds we use the Weibull distribution, which allows us to control the amount of strength disorder of nodes by tuning a single parameter $m$, i.e.\ the Weibull exponent (for details see Methods). 
After a node fails, its load is equally redistributed through the network of load transmitting connections to its nearest neighbors which remained intact during the failure process. As a consequence of this localized load sharing (LLS) the load of neighboring nodes may exceed their local strength resulting in secondary failure which in turn is followed again by load redistribution. Through such sequences of failure and load redistribution steps, a single failing node may trigger an entire cascade of failure events which either stops after a finite number of steps or destroys the entire system. Of course, the statistical and dynamical features of the failure process depend on both the way of load sharing and the amount of disorder in the strength of nodes.
Besides localized load sharing through direct links, the opposite limit of equal load sharing (ELS) has also practical importance. Under ELS conditions all intact nodes of the system receive the same load increment irrespective of their distance from the failed one. Since no load fluctuations can arise, ELS realizes the mean field limit of FBMs, where the failure process is controlled by the quenched strength disorder of nodes (see Methods for the details of the mean field solution of the model).

To reveal scaling laws governing cascading failure phenomena on complex networks, we performed computer simulations  varying the size of the system $L$ in a broad range $100\leq L \leq 2000$ at several rewiring probabilities $p$ considering two different values $m=1$ and $m=3$ of the Weibull exponent. 
For more details of the model construction see Methods. In the simulations the external load $\sigma_0$ was slowly increased to initiate the failure of a single node then the emerging avalanche was generated, while keeping the external load fixed. The process was followed until a catastrophic avalanche destroyed the remaining intact nodes defining the critical load $\sigma_c$, i.e.\ the ultimate strength of the system.

\subsection*{Load bearing capacity and damage tolerance}
The disorder of the strength of the microscopic elements of a system plays a crucial role in failure processes. It has been demonstrated for fracture phenomena that local strength fluctuations can lead to early failure nucleation already at low loads reducing the global strength $\sigma_c$ compared to homogeneous systems \cite{alava_statistical_2006,alava_role_2008,alava_size_2009}. 
Additionally, disorder gives rise to sample-to-sample fluctuations of strength $\sigma_c$ with an average value which decreases with the system size \cite{alava_size_2009}. 
This so-called statistical size effect of the ultimate strength has a great importance for applications: on the one hand it has to be taken into account in engineering design of large scale structures, and on the other hand, it controls how results of laboratory measurements or computer simulations can be scaled up to real constructions \cite{alava_statistical_2006,alava_role_2008,alava_size_2009,yamamoto_PhysRevE.83.066108}. 

\begin{figure}[ht]
\begin{center}
\includegraphics[scale=0.65]{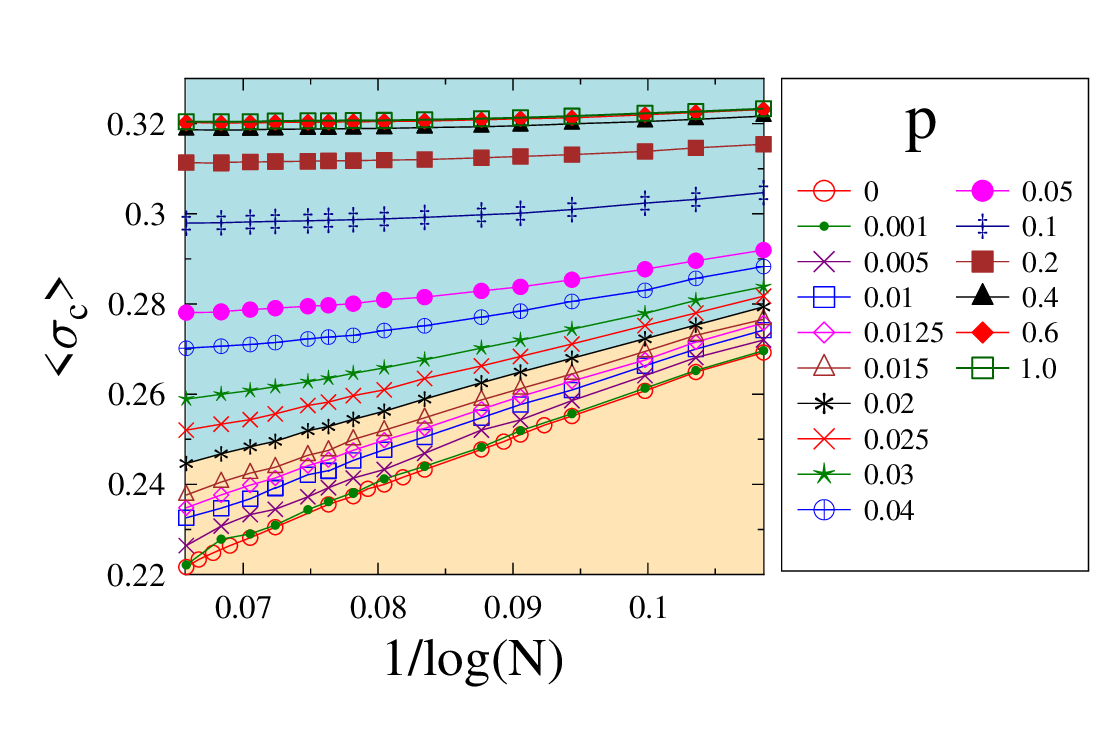}
\caption{The average value of the macroscopic strength $\left<\sigma_c(N,p)\right>$ of finite size bundles as a function of $1/\ln(N)$ in the range of the system size $10^4\leq N \leq 4\times 10^6$ for several values of the rewiring probability $p$. The regimes of $p$ with zero and finite asymptotic strength are highlighted by different background colors. The Weibull exponent $m$ of the threshold distribution is fixed to $m=1$. The value of $p$ increases from bottom to top.}
\label{fig:strength_log}
\end{center}
\end{figure}
To understand how the size scaling of the failure strength is affected by the network structure of load transmitting connections, we evaluated the size dependence 
of the average global strength $\left<\sigma_c(N,p)\right>$ at several rewiring probabilities $p$.  
For ELS, analytic calculations \cite{smith_asymptotic_1982,mccartney_statistical_1983,hansen2015fiber} have revealed that the average failure strength $\left<\sigma_c\right>$ decreases with the number of fibers $N$ and in the limit of large bundle sizes $N$ it converges to a finite value according to a power law 
\beqa{
\left<\sigma_c(N)\right> &=& \sigma_c(\infty) + AN^{-\alpha}, \label{eq:els_sigc}
}
where $\sigma_c(\infty)$ denotes the asymptotic bundle strength.
The scaling exponent $\alpha$ has the value $\alpha^{ELS}=2/3$ which proved to be universal
for a broad class of disorder distributions, while the multiplication factor $A$
and the asymptotic strength $\sigma_c(\infty)$ depend on the specific type of disorder \cite{hansen2015fiber}. 
For the opposite limit of localized load sharing LLS, numerical calculations showed that due to the load concentration around failed regions the system becomes more vulnerable, hence, the macroscopic 
strength of bundles diminishes as the system size $N$
increases. The convergence to zero strength is logarithmically slow with the functional form 
\beq{
\left<\sigma_c\right>(N) \sim 1/(\ln{N})^{\mu},
\label{eq:strengthsize_lls}
}
where the exponent $\mu$ was found to depend on the precise range of load sharing 
\cite{harlow_pure_1985,hansen_burst_1994,hidalgo_fracture_2002,dill-langer_size_2003,
yewande_time_2003,bazant_activation_2007,lehmann_breakdown_2010,pradhan_failure_2010}.
Figure \ref{fig:strength_log} presents the average bundle strength 
$\left<\sigma_c(N,p)\right>$ of our system as a function of $1/\ln{N}$ for several rewiring probabilities $p$ at the Weibull exponent $m=1$.  
It is important to note that for sufficiently low $p$ values the curves tend towards 0 for large $N$ as expected for the LLS phase of failure phenomena.
\begin{figure}[ht]
\begin{center}
\includegraphics[scale=0.61]{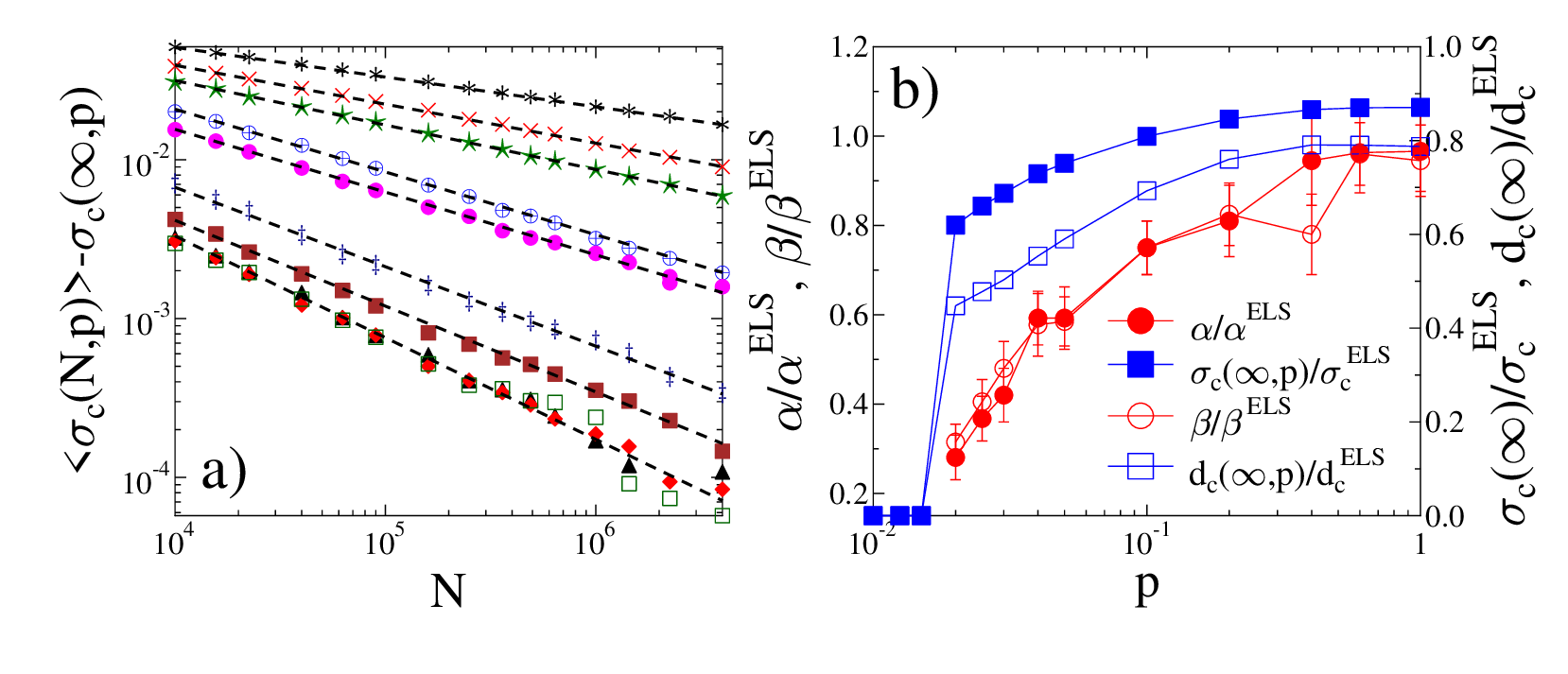}
\caption{$(a)$ Difference of the average strength of finite bundles $\left<\sigma_c(N,p)\right>$ and their corresponding asymptotic strength $\sigma_c(\infty,p)$ for rewiring probabilities in the range $p\geq 0.02$. The dashed straight lines represent power laws of exponent $\alpha$. The legend is the same as in Fig.\ \ref{fig:strength_log}$(a)$. $(b)$ The value of the exponent $\alpha$ and the asymptotic strength $\sigma_c(\infty,p)$ obtained in Fig.\ \ref{fig:sigmacnfit}$(a)$ as function of the rewiring probability $p$. The scaling exponent $\beta$ of the critical damage and the asymptotic damage $d_c(\infty,p)$ are also presented. Note that the parameter values $\alpha$, $\beta$, $\sigma_c(\infty,p)$, and $d_c(\infty,p)$ are rescaled with their mean field counterparts $\alpha^{ELS}$, $\beta^{ELS}$, $\sigma_c^{ELS}$, and $d_c^{ELS}$, respectively. Error bars are provided for the values of the exponents.}
\label{fig:sigmacnfit}
\end{center}
\end{figure}
However, around $p\approx 0.02$ the curvature of the curves changes, indicating the emergence of 
a finite asymptotic strength $\sigma_c(\infty,p)>0$ characteristic for ELS systems. To determine the value of 
$\sigma_c(\infty,p)$ for different network topologies, in Figure \ref{fig:sigmacnfit}$(a)$ we replotted
the $\left<\sigma_c(N,p)\right>$ curves in the parameter range $p \geq 0.02$ by subtracting a proper asymptotic value $\sigma_c(\infty,p)$
in such a way that $\sigma_c(\infty,p)$ was tuned until the best straight line was achieved on a double logarithmic plot. The excellent quality of the power laws in Fig.\ \ref{fig:sigmacnfit}$(a)$ confirms the 
existence of a finite asymptotic strength and the validity of the scaling form Eq.\ (\ref{eq:els_sigc}) of ELS, even in the range of rather low values of the rewiring probability. 
Figure \ref{fig:sigmacnfit}$(b)$ demonstrates that both the asymptotic strength $\sigma_c(\infty,p)$ and the size scaling exponent $\alpha(p)$ depend on the network structure increasing towards their corresponding mean field values $\sigma_c^{ELS}$ and $\alpha^{ELS}$ with increasing rewiring probability $p$. In the localized phase the asymptotic strength is set to be zero in the figure, while no exponents were assigned to this parameter regime.
\cite{smith_asymptotic_1982,mccartney_statistical_1983,hansen2015fiber}. 
Note that in the limit of completely random networks $p\to 1$, both the exponent $\alpha(p=1)$ and the asymptotic strength $\sigma_c(\infty, p=1)$ remain below their corresponding mean field values $\alpha^{ELS}$ and $\sigma_c^{ELS}$ (see Methods for the critical load $\sigma_c^{ELS}$ of ELS FBMs). The reason is that during the rewiring process the average number of interacting partners of nodes remains constant at a relatively low value $\left<k\right>=4$, so that localized load sharing may induce load concentrations even on a random network which can result in global failure at lower loads compared to the mean field limit.

Another important measure of the overall robustness of the system is the amount of damage $d$ the network can tolerate before a catastrophic cascade destroys the entire system. The degree of damage accumulated during the failure process can be quantified by the fraction of failed nodes $d=N_b/N$, where $N_b$ denotes the number of failed nodes among the initially existing $N$ intact ones. The overall damage tolerance of the system is characterized by the critical damage $d_c$ reached up to the last stable configuration
before catastrophic failure occurs. Our numerical analysis revealed that the average critical damage $\left<d_c(N,p)\right>$ exhibits the same qualitative evolution with the rewiring probability $p$, and obeys the same scaling law when the size of the system is varied as the ultimate strength $\left<\sigma_c(N,p)\right>$: at low rewiring probabilities $p\lesssim 0.02$ when the structure of the network is close to regular, the critical damage tends to zero in the limit of large system sizes. This behavior implies that the failure of the weakest node may trigger the immediate catastrophic collapse of the system, as it is expected for the LLS phase of failure phenomena. However, at sufficiently high $p$ values $p > 0.02$ a finite asymptotic damage is obtained according to the functional form similar to Eq.\ (\ref{eq:els_sigc})
\beq{
\left<d_c(N,p)\right> = d_c(\infty,p)+BN^{-\beta}.
\label{eq:scaling_dam}
}
Here $d_c(\infty,p)$ is the critical damage of the infinite system at the rewiring probability $p$, and $\beta$ denotes the scaling exponent. It can be observed in Fig.\ \ref{fig:sigmacnfit}$(b)$ that the value of $\beta$ increases with the rewiring probability $p$ practically coinciding with the exponent $\alpha$ of the ultimate strength. 
The mean field value of the size scaling exponent of critical damage $\beta^{ELS}$ is not known analytically, hence, we determined it by the numerical analysis of direct ELS simulations as $\beta^{ELS}= 0.66\pm0.02$. Based on the numerical results we conjecture that the size scaling of the ultimate strength and of the critical damage is governed by the same exponent $\alpha(p)=\beta(p)$ at all rewiring probabilities outside the localized phase of the failure process. The asymptotic damage $d_c(\infty,p)$ increases with the rewiring probability $p$ towards its mean field value $d_c^{ELS}$, which can be determined analytically (see Methods). Computer simulations revealed that the threshold value of $p$ where the LLS to ELS transition sets on, furthermore, the asymptotic load bearing capacity $\sigma_c(\infty,p)$ and damage tolerance $d_c(\infty,p)$ depend on the disorder distribution $p(\sigma_{th})$ of nodes' strength, however, the finite size scaling exponents $\alpha(p)$ and $\beta(p)$ always cover the same range up to their universal mean field values.

\subsection*{Spatial evolution of damage} 
\begin{figure}[ht]
\begin{center}
\includegraphics[scale=0.6]{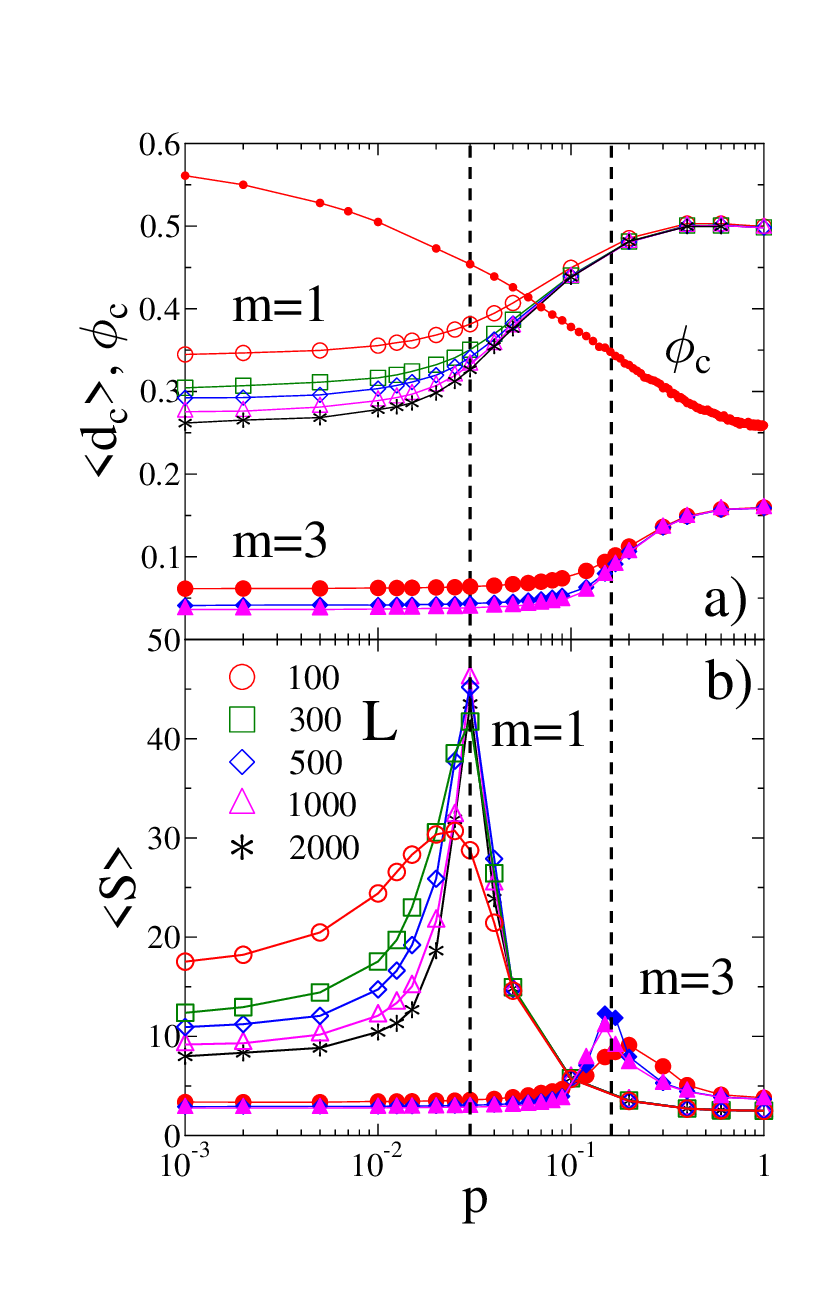}
\caption{$(a)$ The average critical damage $\left<d_c\right>$ $(a)$ and the average size of clusters of failed nodes $\left<S\right>$ $(b)$ as function of the rewiring probability $p$ for several system sizes $L$. The results are presented for two values of the Weibull exponent $m=1$ (open symbols) and $m=3$ (filled symbols). For $m=3$ only three system sizes $L=100, 500, 1000$ are shown. In $(a)$ the critical value of the occupation probability $\phi_c$ of site percolation on the network is also presented as a function of $p$.}
\label{fig:m2m1}
\end{center}
\end{figure}
Nearest neighbor load sharing has the consequence that failed nodes form connected clusters on the network
which have a high load concentration on the intact nodes along their perimeter. Hence, to understand the microscopic origin of the localized to mean field transition and the emergence of scaling laws Eqs.\ (\ref{eq:els_sigc},\ref{eq:scaling_dam}), we analyzed how the spatial structure of damage evolves as the underlying network topology is tuned from regular to completely random.
Since highly loaded nodes are more prone to failure, with increasing external load it gets more and more probable that avalanches are initiated from perimeter nodes giving rise to the growth of already existing clusters instead of nucleating new ones. Hence, clusters are composed of either single cascades (new nucleations) or
of several cascades, which are growth steps of clusters. On a regular lattice ($p=0$) the stress concentration 
is so high at cluster perimeters that the system tolerates only a small amount of damage $d_c\ll 1$ forming 
small clusters randomly dispersed over the network. Introducing long range randomized connections reduces the stress concentration and at the same time it allows small broken clusters, formed on locally regular regions of the network, get connected into larger ones.
As a consequence, increasing the rewiring probability $p$ at a fixed system size $L$ the critical damage $\left<d_c\right>$ increases accompanied by a structural change of failed clusters. This trend of $\left<d_c\right>$ with increasing structural randomness is demonstrated in Fig.\ \ref{fig:m2m1}$(a)$ for two values of the Weibull exponent $m=1$ and $m=3$: for low rewiring probabilities $\left<d_c\right>$ remains nearly constant close to the value obtained on regular lattices $p=0$. At each system size $L$ there exists a characteristic rewiring probability $p\approx 0.02-0.03$ ($m=1$) and $p\approx 0.15-0.2$ ($m=3$), where $\left<d_c\right>$ starts to increase indicating the onset of the transition from the LLS to ELS regime, where the system can tolerate a higher amount of damage. Note that in the LLS regime the curves are rapidly shifting downward with increasing $L$, in agreement with the size scaling result of $\left<d_c(N,p)\right>$ presented in the previous section. However, in the ELS regime the damage curves $\left<d_c(N,p)\right>$ converge towards the same mean field limit $d_c^{ELS}$ at all system sizes $L$ as expected. Note that reducing the strength disorder of nodes by increasing the Weibull exponent $m$ in the figure, the damage curves have the same evolution with increasing structural disorder $p$, however, failure occurs at a significantly lower amount of damage and the LLS to ELS transition sets on at a higher rewiring probability $p$. The results are in a good agreement with recent findings on the effect of disorder on the localized to mean field transition on complex networks \cite{attia_chaossolit_2022, attia_chaos_2022}.

Our calculations revealed that behind this smooth monotonous increase of the accumulated damage $\left<d_c\right>$ with the rewiring probability $p$, a complex evolution of the spatial structure of failed clusters occurs. 
To characterize this evolution, in the last stable configuration of the network we identified all clusters of failed nodes and determined their average size $\left<S\right>$ defined as the average of the ratio of the second $M_2$ and first $M_1$ moments of cluster sizes $S_i$
\beq{
\left<S\right> =\left<M_2/M_1\right>.
}
The $q$th moments $M_q$ with $q=1, 2$ of the size of failed clusters were calculated in single samples as $M_q=\sum^{\prime} S_i^q$, where the $\prime$ indicates that the largest cluster of size $S_{max}$ is always skipped in the summation \cite{stauffer_introduction_1992}. 
\begin{figure}[ht]
\begin{center}
\includegraphics[scale=0.6]{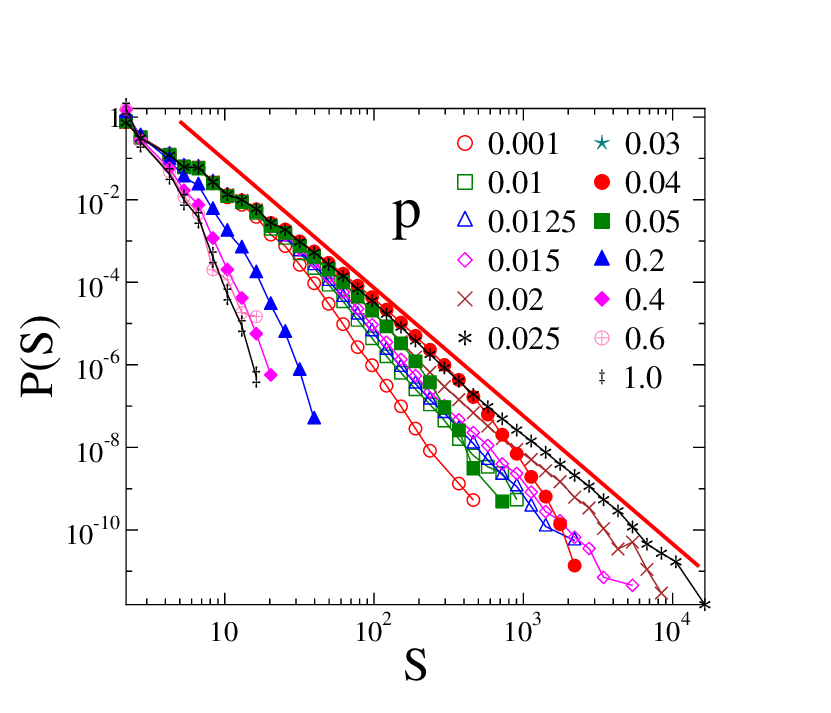}
\caption{Size distribution of clusters of failed nodes $p(S)$ at several rewiring probabilities $p$ for the system size $L=2000$. The Weibull exponent was set to $m=1$. The straight line represents a power law of exponent $\tau=3.1$.}
\label{fig:clustsizedist}
\end{center}
\end{figure}
Figure \ref{fig:m2m1}$(b)$ demonstrates that the average cluster size $\left<S(L,p)\right>$ exhibits a non-monotonous behavior, i.e.\ for each system size $L$ the $\left<S(L,p)\right>$ curves have a relatively sharp peak which gets narrower and slightly higher approaching a limit curve as the system size $L$ increases. As clusters grow to larger sizes with increasing $p$ they have an increasing chance to merge through random long range contacts without destabilizing the system. Since the largest cluster was omitted in the calculations, the position of the peak of $\left<S(L,p)\right>$ marks that network structure, where a dominating cluster of failed nodes, significantly larger than the other ones, first emerges in the last stable configuration of the network before the catastrophic avalanche \cite{stauffer_introduction_1992}. It is important to emphasize that the position of the maximum practically coincides with the rewiring probability $p$, where the LLS to ELS transition sets on as the network gets gradually randomized in agreement with Refs.\ \cite{attia_chaossolit_2022,attia_chaos_2022}. This is highlighted in Fig.\ \ref{fig:m2m1} by the two dashed vertical lines for both Weibull exponents $m$.

Of course, the changing spatial structure of clusters affects also the statistics of their size $S$.
Figure \ref{fig:clustsizedist} illustrates for the Weibull exponent $m=1$ how the cluster size distribution $p(S)$ evolves as the rewiring probability is gradually increased at a fixed system size $L=2000$, where the largest cluster of size $S_{max}$ was always omitted. Until $p$ is sufficiently small, i.e.\ in the LLS phase, 
\begin{figure}[ht]
\begin{center}
\includegraphics[scale=0.6]{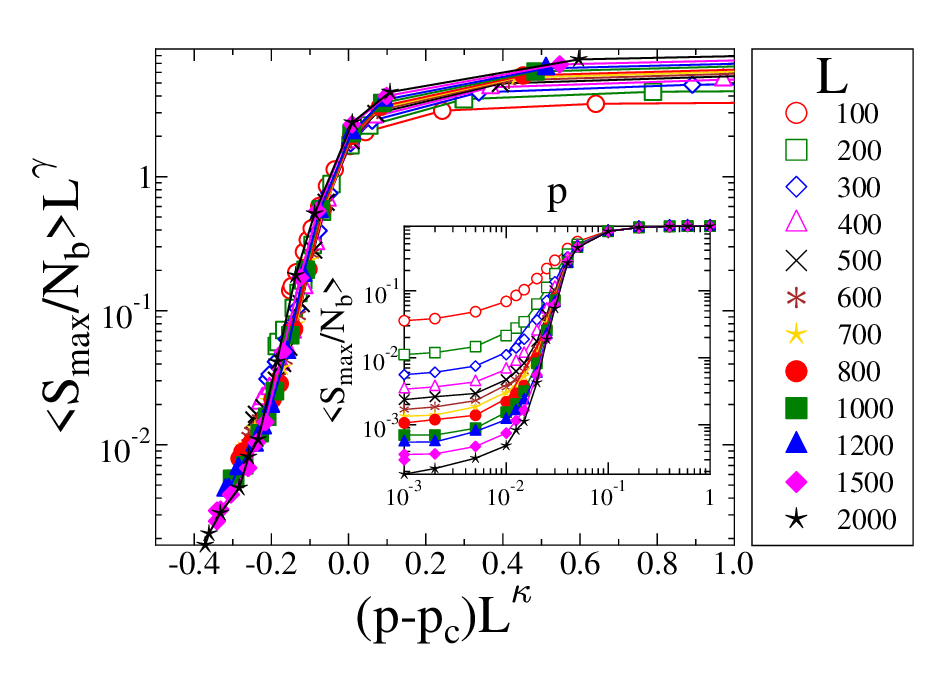}
\caption{Inset: The average strength of the largest cluster $\left<S_{max}/N_b\right>$ as a function of the rewiring probability $p$ for several system sizes at the Weibull exponent $m=1$. Main panel: rescaling the curves of different system sizes of the inset according to the finite size scaling relation Eq.\ (\ref{eq:percordpar}) of the infinite cluster of percolation a good quality data collapse is achieved.}
\label{fig:orderparam}
\end{center}
\end{figure}
a rapidly decreasing functional form is obtained with a relatively low cutoff value. At higher rewiring 
probabilities $p$ the network tolerates larger clusters indicated by the increasing cutoff
and the slower decrease of the distributions $p(S)$. It is interesting to note that at the rewiring probability $p\approx 0.02-0.03$, where the LLS to ELS transition is expected, for the strength disorder $m=1$, the  distribution becomes a power law 
\beq{
p(S) \sim S^{-\tau},
}
spanning three orders of magnitude in $S$ with a relatively high exponent $\tau= 3.1\pm0.15$.
Further increasing the rewiring probability the behavior of the distributions becomes similar to 
the low $p$ case, i.e.\ $p(S)$ is again steeply decreasing limited by a low cutoff. In agreement with the behavior of the average cluster size $\left<S\right>$, the overall evolution of the size distribution of failed clusters indicates that at sufficiently high rewiring probabilities, i.e.\ outside the LLS phase, the largest cluster which was omitted in the statistics comprises a dominating fraction of failed nodes so that even the second largest cluster is significantly smaller than the largest one. The dominating cluster appears at a well defined critical value of the rewiring probability, where the distribution $p(S)$ becomes a power law and the average cluster size $\left<S\right>$ has its maximum accompanied by the onset of the faster increase of the critical damage $\left<d_c\right>$ (see Figs.\ \ref{fig:m2m1},\ref{fig:clustsizedist}). Due to the relatively low value of $p$, locally the structure of the network is still close to regular, however, the long range random connections already allow for the merging of small clusters into a dominating one in the last stable configuration of the network. Note that for those rewiring probabilities which fall beyond the maximum of the average cluster size $\left<S\right>$, the dominating cluster occurs earlier and earlier with increasing $p$ before catastrophic failure so that it has a stronger and stronger effect on the failure process as the external load increases.

\begin{figure}[ht]
\begin{center}
\includegraphics[scale=0.33]{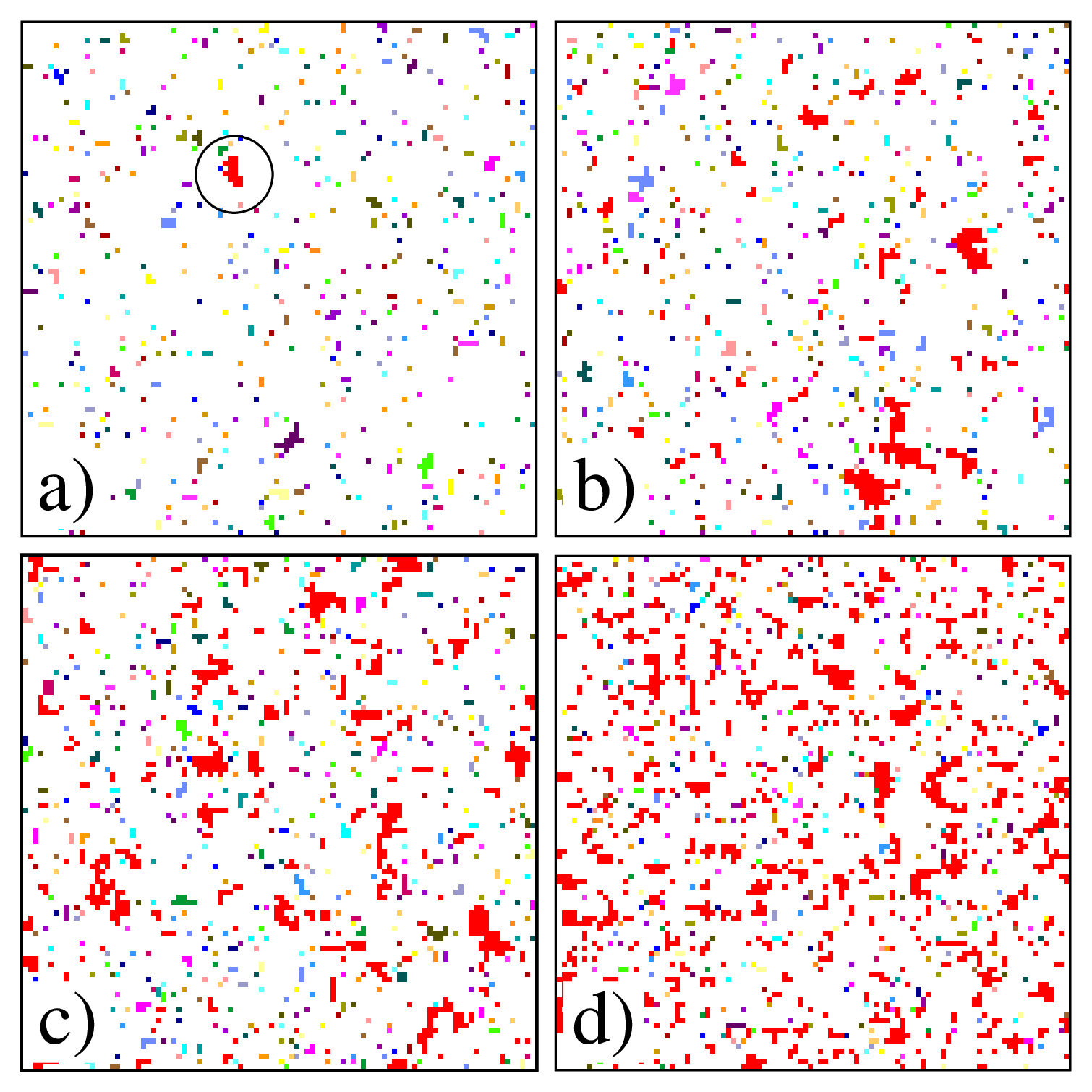}
\caption{Spatial structure of damage on rewired square lattices of size $L=100$ for the Weibull exponent $m=3$. Clusters of failed nodes are presented in the last stable configuration of the system before final catastrophic collapse at four different rewiring probabilities $p$: $(a)$ 0.05, $(b)$ 0.175, $(c)$ 0.225, and $(d)$ 0.4. Colors are randomly assigned to clusters of failed nodes, while the intact nodes are white. The largest cluster is highlighted in red. For clarity, in $(a)$ the largest cluster is encircled.}
\label{fig:snapshots}
\end{center}
\end{figure}
The qualitative behavior of the average cluster size $\left<S\right>$ and of the cluster size distribution $p(S)$ implies that the evolution of the spatial structure of accumulating damage with the changing network topology shows interesting resemblance to percolation phenomena. To obtain a deeper understanding of the analogy of the LLS to ELS transition of failure phenomena to the percolation transition, we analyzed the behavior of the largest cluster of damage. For this purpose, we determined the average value of the strength of the largest cluster $\left<S_{max}/N_b\right>$, defined as the fraction of failed nodes comprised by the largest cluster at the critical point of global failure, as a function of the rewiring probability $p$ at different system sizes $L$. The inset of Fig.\ \ref{fig:orderparam} demonstrates that at each system size $L$ the $\left<S_{max}/N_b\right>$ curves monotonically increase, however, they become steeper and approach a limit curve with growing $L$. The main panel of the figure demonstrates that rescaling the two axis using the finite size scaling relation of the infinite cluster of percolation phenomena \cite{stauffer_introduction_1992}
\beq{
\left<S_{max}/N_b\right>(L,p) = L^{-\gamma}\Phi((p-p_c)L^{\kappa}),
\label{eq:percordpar}
}
the curves of different system sizes $L$ can be collapsed on the top of each other. 
Here $\Phi(x)$ denotes the scaling function obtained by the data collapse analysis in Fig.\ \ref{fig:orderparam}. Best collapse is achieved with the scaling exponents $\kappa=0.32$ and $\gamma=0.35$ using the parameter value $p_c=0.037$. 
The validity of the scaling ansatz Eq.\ (\ref{eq:percordpar}) and the high quality of the data collapse in Fig.\ \ref{fig:orderparam} suggest that the structural transition observed in the accumulated damage, as the network topology changes, shows analogy to a site percolation process on the load transmission network. 

Percolation has been widely studied on complex networks in the context of the robustness of the system against random attacks where nodes of the network are randomly removed (an active node is randomly replaced by a failed one)
\cite{scalperc_newman_pre1999,epiperc_newman_pre2000,newman_epidemic_pre2002,barabasi_network_2016}.
Increasing the fraction $\phi$ of removed nodes, at a certain value the active part of the system falls apart into a large number of small clusters and the system looses its functionality. Failed nodes also form clusters on the network such that at the critical point $\phi_c$ a giant cluster of failed elements emerges. To understand the analogies to site percolation we numerically measured the critical fraction $\phi_c$ of randomly removed, i.e.\ failed nodes where a giant failed cluster emerges on the load transmitting network obtained by rewiring the square lattice at different values of the rewiring probability $p$. It can be observed in Fig.\ \ref{fig:m2m1}$(a)$ that $\phi_c$ is a monotonically decreasing function of $p$ starting from the well known critical point $\phi_c\approx 0.5923$ of site percolation on a square lattice $p=0$ \cite{stauffer_introduction_1992}, and converging to the critical occupation probability of random graphs $\phi_c\approx 1/\left<k\right>=1/4$ in the limit $p\to 1$ \cite{boccaletti_complex_2006}. In our FBM on complex networks the fraction of failed nodes $d$ plays the role of the occupation probability $\phi$ of the percolation problem. It is important to note that at the onset of the LLS to ELS transition $p\approx 0.02-0.03$ at the Weibull exponent $m=1$ in Fig.\ \ref{fig:m2m1}$(a)$ the value of the critical damage does not reach the corresponding critical occupation probability $\phi_c(p)$ of site percolation. The reason is that in our system the load bearing capacity of the bundle is only lost when a catastrophic cascade destroys all the remaining intact fibers so that global failure can also be initiated in the absence of a giant failed cluster as it has been presented in Fig.\ \ref{fig:m2m1}$(b)$. Comparing $\phi_c(p)$ to the $\left<d_c(p)\right>$ curves obtained at different Weibull exponents, it is clear that the difference of the two quantities is even higher at lower amount of strength disorder of nodes (at higher $m$). At the higher Weibull exponent $m=3$ already a relatively small amount of damage $\left<d\right>\approx 0.07$ can trigger ultimate failure of the system on a regular lattice $p\approx 0$ (see Fig.\ \ref{fig:m2m1}$(a)$). The system is so sensitive to load concentrations around failed clusters on the network that even in the limit $p\to 1$ the value of critical damage $\left<d_c\right>$ remains below the corresponding critical point of site percolation $\phi_c$. The results indicate that in spite of the relatively small damage $\left<d_c\right>$ compared to the corresponding value of $\phi_c$, the evolution of the failed cluster, which dominates damage growth, with the rewiring probability is similar to the behavior of the infinite cluster of percolation phenomena. After this dominating cluster has formed it has a substantial effect on the further damage accumulation reducing the load concentration which in turn makes the failure process similar to the ELS limit. It can be observed in Fig.\ \ref{fig:m2m1}$(b)$ that at lower disorder the height of the peak of $\left<S\right>$ gets lower which indicates that the largest cluster comprises a less significant fraction of damage, and hence, has a diminishing effect on the failure process. The result implies that the localized to mean field transition is limited to high disorder in agreement with Refs.\ \cite{attia_chaossolit_2022,attia_chaos_2022}.


\section*{Discussion}
Cascading failure driven by the redistribution of load over the elements of a complex system is strongly affected by the topology of the underlying network of load transmitting connections which shows up both on the macro- and micro-scale characteristics of the system. It has been shown that gradually randomizing an initially regular lattice of connections a transition occurs from the localized to the mean field universality class of failure phenomena, characterized by the abrupt failure of the system preceded by a small amount of precursors, and by a large amount of precursory failure cascades with a scale free statistics, respectively. Here we performed a computational study to understand how the size scaling of failure characteristics are affected by the network structure, which has a high relevance for applications. In particular, we investigated how the macroscopic load bearing capacity and overall damage tolerance of the system scales with the size of the network as the degree of structural randomness of load transmitting connections is varied from completely regular to completely random, at different amounts of the strength disorder of nodes.  

Based on computer simulations we obtained two distinct scaling regimes of macroscopic quantities, i.e.\ 
at low rewiring probabilities both the ultimate strength and the critical damage were found to tend towards zero in the limit of large system sizes, consistent with the localized universality class of failure phenomena. Simulations revealed that as the transition to the mean field class sets on with increasing structural randomness, a finite asymptotic strength emerges which gradually grows towards the corresponding mean field value. Based on the numerical results, we conjectured that the convergence of the ultimate strength and damage tolerance of the system towards their finite asymptotic values is described by the same power law functional form controlled by the same value of the scaling exponent at any network topology. 

The localized redistribution of load through the links of the load transmitting network has the consequence that failed nodes form connected clusters. Computer simulations revealed that increasing the rewiring probability at a fixed system size the network can tolerate a larger amount of damage accompanied by a change of the spatial structure of failed clusters. Inside the localized phase of the system failed clusters remain small even in the last stable configuration just before the catastrophic avalanche is initiated. Randomization of the network structure introduces long range links which make possible the merging of clusters, hence, reducing the load concentration along their perimeter. Consequently, at higher structural randomness networks can tolerate a higher amount of damage before catastrophic failure occurs. 
Most notably, we demonstrated that the localized to mean field transition of the failure dynamics on the micro-level is accompanied by a structural transition of damage on the network, which shows analogies to percolation \cite{stauffer_introduction_1992} up to some extent. Our analysis suggests that $p_c$ obtained from the data collapse analysis Eq.\ (\ref{eq:percordpar}) of the strength of the largest failed cluster can be identified with the critical rewiring probability of the LLS to ELS transition of the fiber bundle model in the limit of large system sizes. 
The transition sets on at the rewiring probability $p_c$, where a damage cluster emerges which is significantly larger than the other ones. The value of $p_c$ obtained from the evolution of the damage structure is consistent with the boundary of the different scaling regimes of the ultimate strength and total damage within the precision of the calculations. The finite size critical point $p_c(N)$ where the localized to mean field transition sets on can be identified with the rewiring probability $p$ of the position of the maximum of $\left<S\right>$.  

The structural transition of damage is illustrated in Fig.\ \ref{fig:snapshots} for a small system of size $L=100$ with the Weibull exponent $m=3$ of strength disorder, where the dominating damage cluster first occurs in Fig.\ \ref{fig:snapshots}$(b)$.
Note in the spatial structure of damage that small clusters of failed fibers are typically formed on locally regular regions of the network so that the dominating cluster occurs by joining such small sized clusters through long range links. On a regular lattice clusters cannot grow to large sizes since the load accumulating along their perimeter destabilizes them so that the catastrophic avalanche typically starts from a perimeter node. Long range random connections have the important effect that they substantially increase the cluster perimeter reducing the load concentration in the cluster neighborhood. Increasing the rewiring probability $p$ beyond $p_c$ the dominating cluster occurs earlier and earlier before the last stable configuration so that in the parameter regime $p>p_c$ the growth of the largest cluster will dominate the further damage growth till the last stable configuration is reached making the failure process similar to its mean field limit. The LLS to ELS transition is driven by the changing network topology where the rewiring probability plays the role of the control parameter of the transition. The analysis of the damage structure revealed that the LLS-ELS transition is limited to high strength disorder of nodes because at low disorder the largest damage cluster cannot comprise a significant fraction of the total damage, and hence, cannot dominate the furter growth of damage.

The occurrence of the transition characterized by the $\langle S\rangle$ peak and the power law $P(S)$ tail at finite $p$ can be understood if we consider that in the 2D LLS phase the cluster size distributions fall rapidly, possibly with an exponential tail. Thus, there is a 1st order like damage transition or perhaps a mixed-order one, where other quantities can exhibit power-law behaviors \cite{odor_griffiths_prr2021}). Analytically known that in the mean-field limit this kind of hybrid phase transition happens, e.g.\ power law behavior of damage is obtained when approaching the critical load of global failure $d_c-d \propto (\sigma_c-\sigma)^{1/2}$ (see \cite{PhysRevE.98.032117,odor_griffiths_prr2021}). This permits to have non-diverging cluster sizes at the LLS to ELS transition point, smaller than the {\og typical distance of WS: $\xi = 1/(2 p)^{1/2}$ (see \cite{scalperc_newman_pre1999}).}
{\og Thus for WS, for $p > p_c$ $\xi < \xi_c$, the correlation length is limited, unlike in other long
range interaction extensions of the FBM, where power-law
distance decaying links are added \cite{N1,N2,N3,N4}. In the latter cases the interaction length can be arbitrarily long, if the decay of the power-law is sufficiently slow.}

{\kf Recently, it has been demonstrated that LLS to ELS transition of FBMs can also be obtained on regular lattices varying the range of interaction, i.e. the range of load sharing among fibers \cite{N3,hidalgo_fracture_2002}. To control the load sharing range two modelling approaches have been considered on a square lattice: the excess load dropped by the broken fiber was either homogeneously distributed over the intact fibers in square shaped plaquettes of the lattice \cite{N3}, or all the intact fibers of the system received an amount of load decaying according to a power law of the distance measured from the broken one \cite{hidalgo_fracture_2002,N3}. Varying also the amount of disorder of the strength of fibers, a phase diagram of FBMs was constructed with a well-defined phase boundary between the nucleation dominated (LLS type) and diffusive (ELS type) fracture mechanisms \cite{N3}. LLS to ELS transition has also been observed by keeping the range of interaction fixed to nearest neighbors on a square lattice but increasing the dimensionality of the system \cite{hansen_lls_dimension_2015,danku_dim_pre}. }

\section*{Methods}
{\it Disordered strength of nodes:} Based on the concept of the fiber bundle model (FBM) \cite{de_arcangelis_scaling_1989,andersen_tricritical_1997,hansen2015fiber,
kloster_burst_1997,kun_extensions_2006,hidalgo_avalanche_2009} we construct a complex network of nodes with disordered load bearing capacities. Under a slowly increasing external load $\sigma_0$ the nodes fail when the local load on them exceeds their local strength $\sigma_{th}$, which is a random variable sampled from a probability distribution $p(\sigma_{th})$. To be able to control the amount of disorder of nodes' strength, we use the Weibull distribution 
\beq{
p(\sigma_{th}) = m\frac{\sigma_{th}^{m-1}}{\lambda^m}\exp{\left[-\left(\frac{\sigma_{th}}{\lambda}\right)^m\right]},
\label{eq:weibull}
}
which is defined over the range $0\leq \sigma_{th}<+\infty$. The distribution has two parameters $\lambda$ and $m$, where $\lambda$ sets the scale of strength values, while the exponent $m$ controls the variance in such a way that the distribution Eq.\ (\ref{eq:weibull}) gets narrower with increasing $m$. The Weibull distribution is widely used to capture the stochastic failure characteristics of components in modeling fracture phenomena from the nanometer scale to the scale of earthquakes varying the value of the Weibull exponent $m$ in the range $m\geq 1$ \cite{kun_extensions_2006,kun_extensions_2006,hidalgo_avalanche_2009}. 

{\it Localized load sharing on a complex network of load transmitting connections:} When a node fails its load has to be overtaken by the remaining intact nodes through the underlying network of load transmitting connections. To generate the network of connections between nodes we start from a regular square lattice of side length $L$ and use the Watts-Strogatz algorithm to randomize the connections \cite{watts_strogatz_nature_1998,watts_networks_1999}. 
Assuming periodic boundary condition in both directions, each link of the square lattice gets rewired with probability $p$: for both ends of a rewired link new nodes are randomly selected with the constraints that neither multiple links nor loops are allowed to occur in the network. Varying the rewiring probability $p$ from 0 to 1 the topology of the load transmission network changes from regular to completely random. The degree $k$ of a node is defined as the number of its direct neighbors along the links of the network. Initially, each node has the same degree $k=4$, due to rewiring the degree distribution $\rho(k)$ broadens while keeping the average node degree fixed $\left<k\right>=4$. 

{\it Cascading failure triggered by external load increments:} To ensure quasi-static loading, the external load is incremented in small steps on the system to provoke the failure of a single node, which is then followed by the redistribution of load. During the failure process we apply localized load sharing (LLS), i.e.\ when a node fails its load is equally redistributed over its intact nearest neighbors through the links of the load transmission network. As a consequence of load redistribution some neighboring nodes may fail followed again by load sharing. Hence, through the consecutive failure-load redistribution steps the failure of a single node may induce an entire cascade of local failure events on the network. The external load $\sigma_0$ was increased until a catastrophic avalanche was triggered destroying the entire system at the critical load $\sigma_c$. In the present study simulations were performed at several values of the rewiring probability $p$ between $0$ and $1$ varying the system size $L$ in the range $100\leq L \leq 2000$. The scale parameter of the Weibull distribution was fixed to $\lambda=1$, while for the exponent $m$ two different values $m=1$ and $m=3$ were considered. The number of nodes $N=L^2=10^4-4\cdot10^6$ proved to be sufficient to deduce the size dependence of key quantities of the system. At each parameter set $K=1000$ samples were simulated for averaging.

{\it Mean field limit of the model:} Another limiting case of load redistribution is the so-called equal load sharing (ELS), where after failure events each intact node of the system receives the same load increment irrespective of its distance from the failed one. Since no load fluctuations can arise, ELS realizes the mean field limit of FBMs. Under ELS conditions important characteristic quantities of the system can be cast into analytical forms. The total load $\sigma_0$ on the system can be expressed as a function of the load $\sigma$ of single nodes as 
\beq{
\sigma_0 = \sigma \left[1-P\left(\sigma\right)\right],
}
where $P(x)$ denotes the cumulative distribution of failure thresholds. The term $\left[1-P\left(\sigma\right)\right]$ provides the fraction of intact nodes of the network when they all keep the same load $\sigma$ during the loading process. For disorder distributions $P(x)$ relevant for practical purposes, the curve of $\sigma_0(\sigma)$ has a maximum whose value provides the  critical load bearing capacity $\sigma_c$ of the network, where the catastrophic cascade occurs. Substituting the Weibull distribution Eq.\ (\ref{eq:weibull}), the critical load $\sigma_c^{ELS}$ of the ELS network follows as  \cite{hansen2015fiber}
\beq{
\sigma_c^{ELS} = \lambda\left(\frac{1}{m}\right)^{1/m}e^{-1/m},
}
while the critical damage $d_c^{ELS}$, i.e.\ the fraction of failed nodes at the instant of global failure, takes the form \cite{hansen2015fiber}
\beq{
d_c^{ELS} = 1-e^{-1/m}.
}
We use the ELS values of the ultimate strength $\sigma_c^{ELS}$ and damage tolerance $d_c^{ELS}$ for comparison to quantify how close the behavior of the network is to the mean field limit.

\section*{Data availability}
Numerical results of computer simulations are available from the corresponding author upon reasonable request.

\section*{Code availability}
The numerical code used for data evaluation in this paper is available from the corresponding author upon reasonable request.

\bibliography{statphys_fracture}

\section*{Acknowledgments}
This research used computational resources of the supercomputer Fugaku provided by the RIKEN Center for Computational Science. The work was supported by the EFOP-3.6.1-16-2016-00022 project. This research was supported by the National Research, Development and Innovation Fund of Hungary, financed under the Projects K 119967 and K 128989. Project no. TKP2020-NKA-04 has been implemented with the support provided from the National Research, Development and Innovation Fund of Hungary, financed under the 2020–4.1.1-TKP2020 funding scheme. 

\section*{Author contributions statement}
GP, ZsD, AB, VK, and NY developed simulation codes, carried out computer simulations, and performed data evaluation. FK, NI, and GO performed and discussed data analysis, conceived and designed the study. FK drafted the manuscript. All authors read and approved the manuscript.


\section*{Competing interests} 
The authors declare no competing financial interests.

\noindent The corresponding author is responsible for submitting a \href{http://www.nature.com/srep/policies/index.html#competing}{competing interests statement} on behalf of all authors of the paper.

\end{document}